\documentclass[aps,twocolumn,preprintnumbers,showpacs,showkeys,nofootinbib,
superscriptaddress]{revtex4}
\usepackage{epsfig}
\usepackage{amssymb,amsmath,amsfonts,amsthm,graphicx,psfrag}
\graphicspath{{./Fig/}}                 % location for color  fig.
\newcommand{\chpt}{\ensuremath{\chi\text{pt}}}

\newcommand{\psibar}{\ensuremath{\overline\psi}}
\newcommand{\chibar}{\ensuremath{\overline\chi}}
\newcommand{\pbp}{\ensuremath{\psibar\psi}}
\newcommand{\pbpvac}{\ensuremath{\langle \psibar\psi \rangle}}
\newcommand{\Nt}{\ensuremath{N_\tau}}
\newcommand{\Ns}{\ensuremath{N_\sigma}}
\newcommand{\vev}[1]{\ensuremath{\left<#1\right>}}
\newcommand{\Real}{\ensuremath{\text{Re}}}
\newcommand{\Trace}{\ensuremath{\text{Tr}}}
\newcommand{\ord}[1]{\ensuremath{\mathcal{O}(#1)}}

\newcommand{\ii}{\ensuremath{\text{i}}}
\newcommand{\eq}[1]{eq.\,(\ref{#1})}
\newcommand{\Eq}[1]{Eq.\,(\ref{#1})}

\newcommand{\Fig}[1]{Fig.\,\ref{#1}}
\newcommand{\Figs}{Figs.\,}

\newcommand{\tab}[1]{table\,\ref{#1}}
\newcommand{\Tab}[1]{Table\,\ref{#1}}

\def\lsi{\raise0.3ex\hbox{$<$\kern-0.75em\raise-1.1ex\hbox{$\sim$}}}
\def\gsi{\raise0.3ex\hbox{$>$\kern-0.75em\raise-1.1ex\hbox{$\sim$}}}
\newcommand{\lsim}{\mathop{\lsi}}

\newlength{\graphicswidth}
\begin{document}
\preprint{HU-EP-11/10, SFB/CPP-11-06 (revised version)}

\title{The thermal QCD transition with two flavors of twisted mass fermions}
%---------------------------------------------------------------------------

\author{Florian Burger}
\affiliation{Humboldt-Universit\"at zu Berlin, Institut f\"ur Physik, 12489 Berlin, Germany}
\author{Ernst-Michael Ilgenfritz}
\affiliation{Humboldt-Universit\"at zu Berlin, Institut f\"ur Physik, 12489 Berlin, Germany}
\affiliation{Joint Institute for Nuclear Research, VBLHEP, 141980 Dubna, Russia} 
\author{Malik Kirchner}
\affiliation{Humboldt-Universit\"at zu Berlin, Institut f\"ur Physik, 12489 Berlin, Germany}
\author{Maria Paola Lombardo}
\affiliation{Laboratori Nazionali di Frascati, INFN, 100044 Frascati, Roma, Italy}
\author{Michael M{\"u}ller-Preussker} 
\affiliation{Humboldt-Universit\"at zu Berlin, Institut f\"ur Physik, 12489 Berlin, Germany}
\author{Owe Philipsen}
\affiliation{Goethe-Universit\"at Frankfurt, Institut f\"ur Theoretische Physik, 
 60438 Frankfurt am Main, Germany}
\author{Christopher Pinke}
\affiliation{Goethe-Universit\"at Frankfurt, Institut f\"ur Theoretische Physik, 
 60438 Frankfurt am Main, Germany}
\author{Carsten Urbach}
\affiliation{Universit\"at Bonn, HISKP and Bethe Center for Theoretical Physics, 
 53115 Bonn, Germany}
\author{Lars Zeidlewicz}
\affiliation{Goethe-Universit\"at Frankfurt, Institut f\"ur Theoretische Physik, 
 60438 Frankfurt am Main, Germany}
\collaboration{tmfT collaboration}
\noaffiliation

\date{December 3, 2012}
%-------------------------------------------

\begin{abstract}
%---------------
We investigate the thermal QCD transition with two flavors of maximally twisted 
mass fermions for a set of pion masses, 300~MeV \textless $m_\pi$ \textless 500~MeV,
and lattice spacings $a$ \textless 0.09~fm. We determine the pseudo-critical temperatures
and discuss their extrapolation to the chiral limit using scaling forms for different 
universality classes, as well as the scaling form for the magnetic equation of state.  
For all pion masses considered we find resonable consistency with $O(4)$ scaling plus 
leading corrections. However, a true distinction between the $O(4)$ scenario and a first 
order scenario in the chiral limit requires lighter pions than are currently in use in 
simulations of Wilson fermions. 
\end{abstract}

\maketitle

\section{Introduction}
%--------------------
The transition from a confined phase with broken chiral symmetry to a deconfined chirally 
symmetric phase is an important subject for studies of finite temperature quantum 
chromodynamics (QCD). This transition is relevant for the evolution of the early universe 
and reproduced in current heavy ion collision experiments. It can be investigated 
non-perturbatively using lattice QCD as long as the chemical potential for fermion number 
is small, $\mu/T<1$. 
A lot of effort has been invested in lattice studies at zero chemical potential, 
for recent reviews see~\cite{Kanaya:2010qj,Levkova:2012jd,Philipsen:2012nu,Lombardo:2012}.
Impressive progress has been reported very recently by several collaborations 
working with different fermion discretization schemes 
\cite{Bazavov:2012jq,Bazavov:2012jaa,Borsanyi:2012xf,Borsanyi:2012uq,
Umeda:2012er,Ejiri:2012vy}.
In particular, lattice QCD with staggered fermions and physical quark masses does not 
predict a true phase transition but an analytic crossover in the limit of zero chemical 
potential~\cite{Aoki:2006we}. Similarly, results on the transition temperature and the equation 
of state have predominantly been obtained from simulations with staggered 
fermions~\cite{Borsanyi:2010bp,Borsanyi:2010cj,Cheng:2006qk,Cheng:2009zi,Bazavov:2011nk}. 
However, this fermion discretization is subject to an on-going debate and there is no 
formal proof that its continuum limit will reproduce the universality class of 
QCD~\cite{Creutz:2008nk}. 
It is therefore desirable to obtain independent results with other discretizations, 
in order to have some mutual control over systematic errors.

Unfortunately, Wilson-type fermions (and even more so chiral fermion formulations) 
require higher computational costs. It is thus expedient to study the nature of the phase 
transition for various larger than physical quark masses and to extrapolate to the physical
situation.  Moreover, knowing global properties of the phase transition as a function of 
the light quark masses constrains the enlarged phase diagram including the strange quark 
and non-vanishing chemical potential~\cite{Philipsen:2009dn}. An as yet unsettled crucial question 
in this context is the nature of the phase transition in the two-flavor chiral limit. Most 
studies favor a second order transition in the $O(4)$ universality 
class~\cite{Karsch:1994hm,Bernard:1996iz,Iwasaki:1996ya,Aoki:1998wg,AliKhan:2000iz,Ejiri:2009ac} 
but there are also claims for a first order transition~\cite{D'Elia:2005bv,Cossu:2007mn,
Bonati:2009yg,Aoki:2012yj,Bonati:2012pe}. Since in the continuum and chiral limits the 
transition is associated with the breaking of a global chiral symmetry,
it is necessarily a true and non-analytic phase transition and one of these scenarios has to be 
realized~\cite{Pisarski:1983ms}, while an analytic crossover is ruled out. On the other hand, for 
moderate and intermediate quark masses, the transition is an analytic crossover, before it 
turns into a first-order deconfinement transition for very heavy quarks.

In this article we study the thermal transition with two degenerate flavors of maximally 
twisted mass fermions, which provide an \ord{a}-improved Wilson fermion discretization, 
for a review see~\cite{Shindler:2007vp}. As a first step we focus on the determination of the 
phase boundary, i.\,e.~the pseudo-critical temperatures $T_c(m_\pi)$ using the Polyakov loop, 
the chiral condensate and the plaquette as observables. We do this for a set of pion masses,  
$m_\pi\approx 300-500$ MeV, and attempt various extrapolations to the $N_f=2$ chiral limit. 
Similar efforts were recently under way employing clover improved fermions
~\cite{Umeda:2008bd,Bornyakov:2011yb,Brandt:2012sk}. 

The following section serves to specify our simulation setup. In Section~\ref{sec:temperature} 
we introduce the observables and collect the pseudo-critical couplings from our simulations. 
These results allow for an estimate of the size of the discretization errors present in our 
simulations. In Section~\ref{sec:chiral} we use these pseudo-critical points for an 
extrapolation to the chiral limit. We discuss possibilities and limitations in discerning 
the order of the chiral phase transition. Finally Section~\ref{sec:concl} gives some conclusions 
and an outlook.

\section{Simulation setup}
%-------------------------
We consider QCD with a mass-degenerate doublet of twisted mass fermions, cf.\ the review by 
Shindler~\cite{Shindler:2007vp}. The gauge action is tree-level Symanzik improved while the 
fermion action is 
\begin{equation}
S_F[U,\psi,\psibar] = \sum_x \chibar(x)\left( 1 -\kappa D_W[U] + 
                    2\ii\kappa a\mu_0 \gamma_5\tau^3\right)\chi(x)\;.
\end{equation}
The fermion fields are written in the twisted basis $\{\chibar,\chi\}$ which is commonly used 
for numerical simulations. It is connected to the basis of physical fields $\{\psibar,\psi\}$ 
for the relevant case of maximal twist via
\begin{equation}
\psi = \frac{1}{\sqrt{2}}(1+\ii\gamma_5\tau^3)\chi \quad \text{and} \quad 
\psibar = \chibar \frac{1}{\sqrt{2}}(1+\ii\gamma_5\tau^3)\;.
\end{equation}
The quark mass is determined by the hopping parameter $\kappa$, which parameterizes the 
untwisted bare quark mass component, 
\begin{equation}
\kappa = (2am_0 + 8r)^{-1}\;,
\end{equation}
and the twisted mass parameter $\mu_0$.
The Wilson covariant derivative is given by
\begin{equation}
\begin{split}
D_W[U]\psi(x) = \sum_\mu \left( \left(r-\gamma_\mu\right) U_\mu(x)\psi(x+\hat\mu) \right. 
\\ \left.+ \left(r+\gamma_\mu\right) U_\mu^\dagger(x-\hat\mu)\psi(x-\hat\mu)\right)\;. 
\end{split}
\end{equation}
In the weak coupling limit, $\beta=6/g_0^2\rightarrow\infty$, zero quark mass corresponds to 
$\kappa=1/8$, setting $r=1$. For finite coupling this value of $\kappa$ gets corrections 
through mass renormalization. The overall renormalized quark mass $M$ is composed of the 
twisted and untwisted masses as
\begin{equation}
M^2 = Z_m^2\left(m_0-m_\text{cr}\right)^2 + Z_\mu^2 \mu_0^2\;.  
\end{equation}

\begin{figure}
\centering
\includegraphics[width=\graphicswidth,angle=-90]{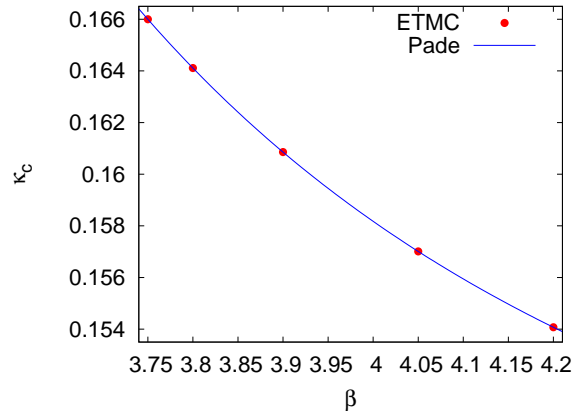}\hfill
\caption{Interpolation of the critical hopping parameter from ETMC data.}
\label{fig:interpol:kc}
\end{figure}

\begin{figure}
\centering
\includegraphics[width=\graphicswidth,angle=-90]{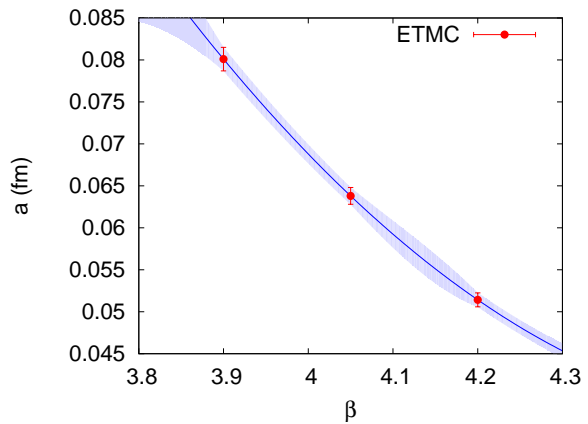}
\caption{Interpolation of the lattice spacing from ETMC data.}
\label{fig:interpol:a}
\end{figure}

At maximal twist, the above fermion formulation is automatically \ord{a}-improved, 
i.\,e.\ cutoff effects linear in the lattice spacing $a$ are absent for non-zero physical 
observables. Maximal twist is achieved by tuning the hopping parameter to its critical 
value $\kappa_c$, corresponding to $m_\text{cr}$, 
where the untwisted theory would feature massless pions. The required knowledge of $\kappa_c(\beta$), 
as well as other input needed from zero temperature simulations in order to set the scale, 
can be interpolated from data by the European Twisted Mass Collaboration (ETMC)~\cite{Baron:2009wt}. 
In \Figs~\ref{fig:interpol:kc} and \ref{fig:interpol:a} we show our interpolations for 
$\kappa_c(\beta)$ and the lattice spacing $a(\beta)$. Our numerical evaluation proceeds 
by an HMC algorithm \cite{Urbach:2005ji} within the publicly available code for QCD with 
twisted mass fermions \cite{Jansen:2009xp}.

Wilson fermions are well-known to feature unphysical phases for light quarks and coarse 
lattice spacings. Like the physical parameter space, these get extended to a third direction
because of the additional twisted mass parameter in the current formulation.
In order to stay away from unphysical regions, knowledge of the bare parameter phase diagram 
is required, which we have mapped out earlier in a preparatory study~\cite{Ilgenfritz:2009ns}.
Status reports of our ongoing project have been given at the annual lattice 
conferences~\cite{Burger:2010ag,MullerPreussker:2009iu}.

The temperature scale is set by the temporal lattice extent and the lattice spacing, 
$T=1/(a\Nt)$. In order to locate the phase boundary between the hadronic region and the quark 
gluon plasma, we perform scans in the lattice gauge coupling $\beta$, which  thus corresponds
to a change in temperature of the lattice system. Table \ref{tab:runs} gives the list of runs 
for different pion masses and the naming scheme that we have adopted for the sake of simplicity. 
To adjust the masses, ETMC provides parameters for NNLO\chpt-formulae at their values of 
$\beta\in\{3.8,3.9,4.05,4.2\}$ which can be used to identify the relation $m_\pi(\mu_0)$ 
at those couplings. For our $\Nt=12$ scans we have relied on the one-loop scaling relation
\begin{equation}
a\mu_0(\beta) = C \exp\left(-\frac{\beta}{12\beta_0}\right)\;,
\end{equation}
with $\beta_0=(11-2N_f/3)/(4\pi)^2$ and fixing the free parameter $C$ at one of the available 
couplings. We have found this relation to work sufficiently well to create lines of constant 
pion mass within the errors quoted in \Tab{tab:runs}. The run at $\Nt=10$ has a constant 
$a\mu_0=0.006$ in the $\beta$-interval from $3.865$ to $3.930$ for which we likewise have the 
same pion mass within errors in our simulation range. 
For the other $\beta$-values we have adapted the twisted mass according to a two-loop scaling 
relation similar to the one-loop formula shown above. The free parameter $C$ has been adapted 
to produce $a\mu_0=0.006$ at $\beta=3.88$.

%-----------------------------------------------------------------------
\begin{table}
\centering
\begin{ruledtabular}
\begin{tabular}{lllll}
\textsc{Run} & $\Ns^3\times\Nt$ & \multicolumn{1}{c}{\textsc{Range}} 
             & $m_\pi$\,(MeV) & \multicolumn{1}{c}{$r_0m_\pi$}\\
\hline
\hline
A12 & $32^3\times12$ & $3.84\le \beta \le 3.99$ & 316(16) & 0.673(42)\\
B12 & $32^3\times12$ & $3.86\le \beta \le 4.35$ & 398(20) & 0.847(53)\\
C12 & $32^3\times12$ & $3.90\le \beta \le 4.07$ & 469(24) & 0.998(62)\\
B10 & $32^3\times10$ & $3.76\le \beta \le 4.35$ & 398(20) & 0.847(53)
\end{tabular}
\end{ruledtabular}
\caption{List of scans in $\beta$. See also \tab{tab:hmcstat}.}
\label{tab:runs}
\end{table}
%------------------------------------------------------------------------

A final comment concerns the explicit flavor symmetry breaking due to the twisted mass term at 
finite values of the lattice spacing. This breaking has been investigated by the ETM collaboration 
for $T=0$ theoretically~\cite{Dimopoulos:2009qv} and in simulations~\cite{Baron:2009wt}. The outcome 
is that effects from flavor breaking -- formally of $O(a^2)$ -- appear to be negligible in all 
quantities investigated so far but the neutral pion mass. For this reason we use the charged 
pion mass throughout the paper. As will be explained in Section \ref{sec:chiral}, for our scaling 
analysis we need to be close enough to the continuum in order to reproduce chiral symmetry,  
where flavor breaking should not play any role any longer. Comparison of two lattice spacings 
appears to justify this assumption. However, a third value of the lattice spacing is required 
in order to make these statements about the size of lattice artifacts more definite.

\section{Thermal transition temperature} \label{sec:temperature}
%---------------------------------------------------------------

In order to locate the transition, we have used both pure gauge and fermionic observables. 
The gauge observables are the plaquette
\begin{equation}
 P = \frac{1}{6N_c \Nt \Ns^3} \Real \Trace \sum_{x}\sum_{\mu>\nu} U_{\mu\nu}(x)\;,
\end{equation}
with
\begin{equation}
U_{\mu\nu}(x) = U_\mu(x)U_\nu(x+\hat\mu)U_\mu^\dagger(x+\hat\nu)U_\nu^\dagger(x)\;,
\end{equation}
and the real part of the Polyakov loop
\begin{equation}
 \Real\left(L\right) = \frac{1}{N_c} \frac{1}{\Ns^3} 
           \Real \Trace \sum_{\mathbf{x}}\prod_{x_4=0}^{\Nt-1} U_4\left(\mathbf{x},x_4\right)\;.
\label{eq:polloop}
\end{equation}
The latter is of particular interest since it is the order parameter of the pure gauge 
deconfinement transition. Along with these observables, we look at their susceptibilities,
\begin{equation}
\chi_O = \Ns^3 \left( \vev{O^2} - \vev{O}^2 \right)\;.
\end{equation}

The renormalized (real part of the) Polyakov loop can be determined as 
\cite{Aoki:2006br}
\begin{equation}
  \vev{\Real(L)}_R = \vev{\Real(L)}  \exp{(V(r_0)/2T)}\,,
\end{equation}
where $V(r_0)$ denotes the static quark-antiquark potential at the 
distance of the Sommer scale $r=r_0$ 
\cite{Sommer:1993ce} 
to be determined at zero temperature. 

The chiral condensate \vev{\pbp} represents the real order parameter of 
chiral symmetry breaking in the massless limit. An appropriate quantity 
to locate the chiral phase transition is the chiral susceptibility
\begin{equation}
\chi_\sigma = \frac{\partial \vev\pbp}{\partial m_q}\;.
\end{equation}
Here, we consider only a part of that expression, the variance per 
configuration, 
\begin{equation}
\sigma^2_{\pbp} = V/T \left( \vev{(\pbp)^2} - \vev{\pbp}^2 \right)\;.
\label{eq:sigma2}
\end{equation}
This quantity shows a peak associated with the chiral transition. 
Moreover, it is expected to dominate the signal of $\chi_\sigma$, 
see e.\,g.~\cite{DeTar:2011nm}.

The pion norm 
\begin{equation}
\vert\pi\vert^2 = \sum_x \vev{\psibar(x)\frac{1}{2}\gamma_5\tau^+
                  \psi(x)\,\psibar(0)\frac{1}{2}\gamma_5\tau^-\psi(0)}
\end{equation}
is interesting for twisted mass simulations because its definition is 
independent of the fermion basis.  It is connected with the chiral condensate via 
\begin{equation}
2m_q \vert\pi\vert^2 = -\vev{\pbp}, \label{eq:ward}
\end{equation}
which has been proven for lattice twisted mass fermions in 
\cite{Frezzotti:2005gi}. We have used this relation as a check for \vev\pbp. 

At maximal twist the chiral condensate can be renormalized as follows 
(see the appendix in \cite{Dinter:2012tt} and references cited therein)
\begin{equation}
\vev{\pbp}_R = Z_P (\vev{\pbp} + c(g_o)\frac{\mu_0}{a^2} ). 
\end{equation}
This immediately suggests the form of a subtracted condensate, which is completely 
standard. However, the subtracted condensate is no longer an order parameter for the 
chiral transition. It is very easy to fix this problem by adding 
the zero temperature chiral condensate in the chiral limit.
Thus, we introduce a (re)normalized condensate in terms of the ratio 
\begin{equation}
  R_{\pbpvac}= \frac{\pbpvac(T,\mu_0) - \pbpvac(0,\mu_0) + \pbpvac(0,0)}{\pbpvac(0,0)}\,,
\label{eq:Rratio} 
\end{equation}
where $\pbpvac(T,\mu_0)$ means $\pbpvac$ to be evaluated at non-zero temperature and 
finite $\mu_0$. $\pbpvac(0,\mu_0)$ can be obtained from spline interpolations of 
$T=0$ $\pbpvac$ data in both the mass $\mu_0$ and $\beta$.
Additionally, to determine $\pbpvac(0,0)$ one has to perform a chiral extrapolation 
of the $T=0$ $\pbpvac$ data at every $\beta$-value one is interested in. 
We have used a linear extrapolation through three points at every $\beta$. 
The data turned out to be compatible with a linear $\mu_0$ dependence over 
the whole temperature range we consider here.  
For the $T=0$ data we were relying on results provided by the ETM collaboration.

The fermionic observables have been determined using the technique of 
noisy estimators, as in \cite{Boucaud:2008xu}. 
For $|\pi|^2$ we have calculated ten propagators per gauge configuration
on $Z(2)$ noise vectors. $\langle \bar \psi \psi \rangle$ was evaluated using 
24 Gaussian volume source vectors for B10, B12 and C12, and 24 $Z(2)$ volume 
source vectors for A12 respectively. 
All propagators have been calculated on commodity graphics hardware using NVIDIA's CUDA 
programming language. The statistics accumulated for the various
runs as well as the averages for the Polyakov loop and the chiral condensate
are given at the end of the paper in \Tab{tab:hmcstat}.

Quite generally, we find the signals for the transition to be 
quite smooth and noisy
which presumably is related to the fact that we are merely probing a 
very soft crossover in our range of pion masses.
For a crossover there is no unique definition of a critical temperature as
the physics changes smoothly and analytically between the different regions.
A pseudo-critical temperature associated with the transition behavior
of individual observables is in general observable-dependent. 

%------------------------------------------------------------------------------
\begin{figure*}
\centering \hfill
\includegraphics[width=\graphicswidth]{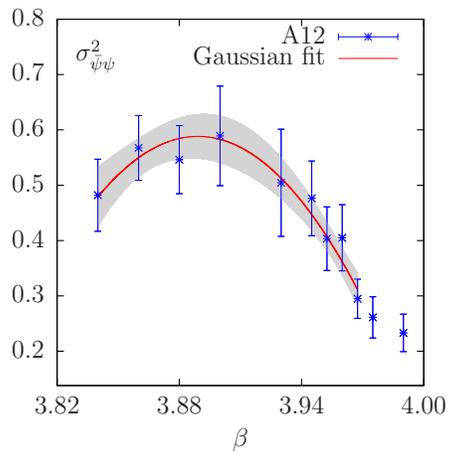} \hfill
\centering \hfill
\includegraphics[width=\graphicswidth]{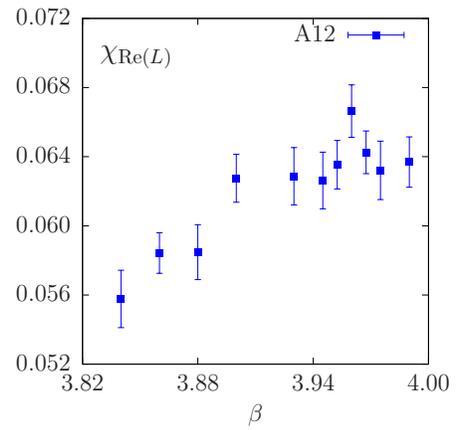}
\caption[]{$\sigma^2_{\pbp}$ (left) and susceptibility of \Real(L)
(right), both for run A12.}
\label{tc_A12}
\end{figure*}
%B12
\begin{figure*}
\centering \hfill
\includegraphics[width=\graphicswidth]{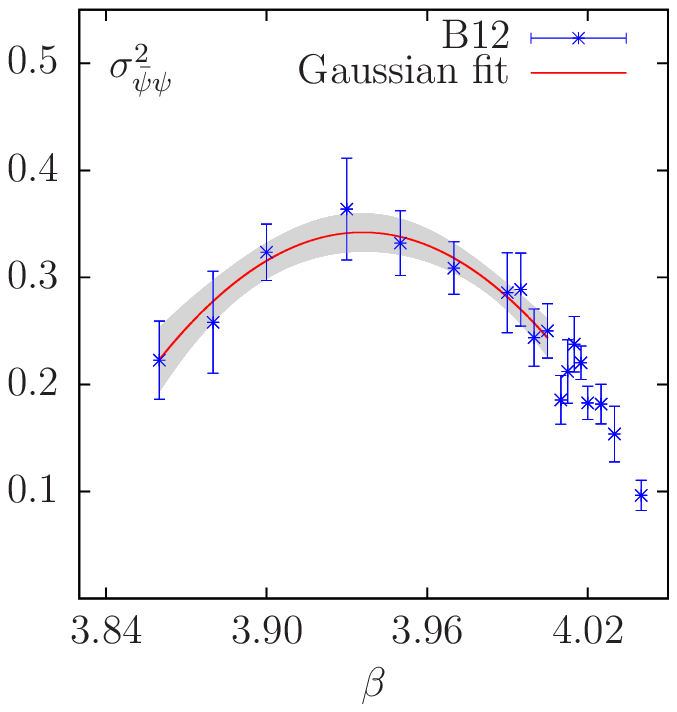} \hfill
\centering \hfill
\includegraphics[width=\graphicswidth]{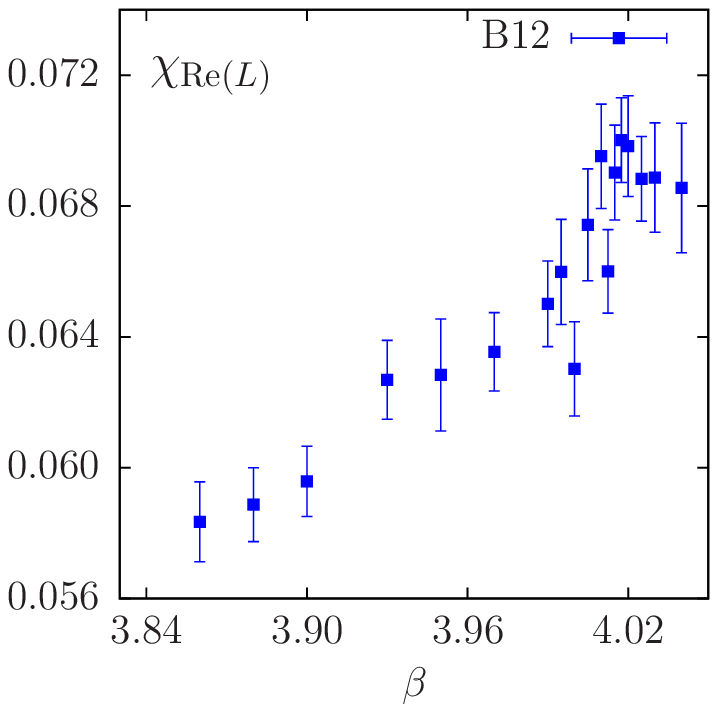}
\caption[]{$\sigma^2_{\pbp}$ (left) and susceptibility of \Real(L)
(right), both for run B12.}
\label{tc_B12}
\end{figure*}
%C12
\begin{figure*}
\centering \hfill
\includegraphics[width=\graphicswidth]{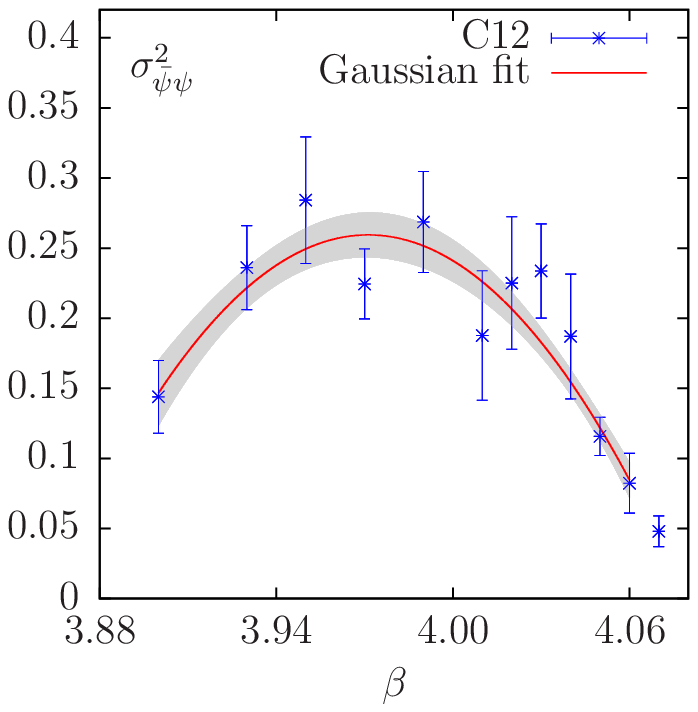} \hfill
\centering \hfill
\includegraphics[width=\graphicswidth]{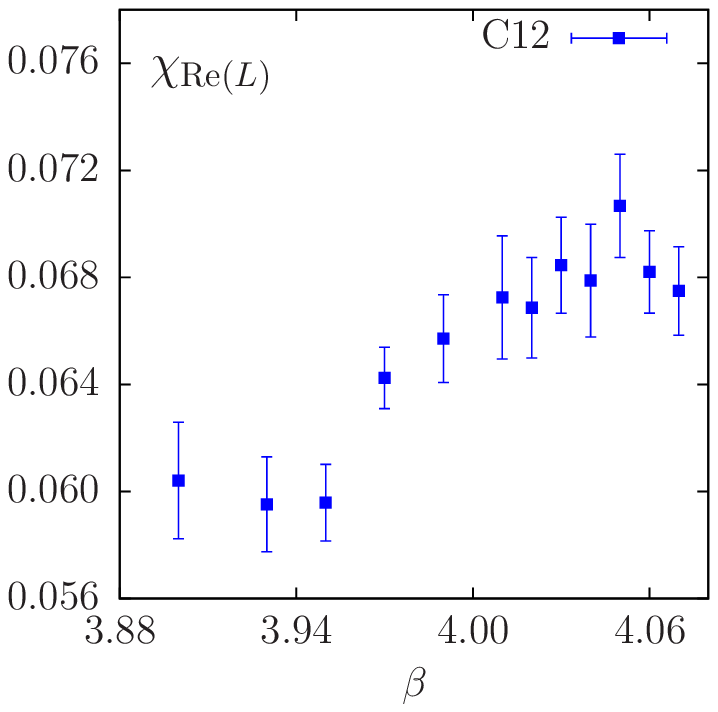}
\caption[]{$\sigma^2_{\pbp}$ (left) and susceptibility of \Real(L)
(right), both for run C12.}
\label{tc_C12}
\end{figure*}
%B10
\begin{figure*}
\centering \hfill
\includegraphics[width=\graphicswidth]{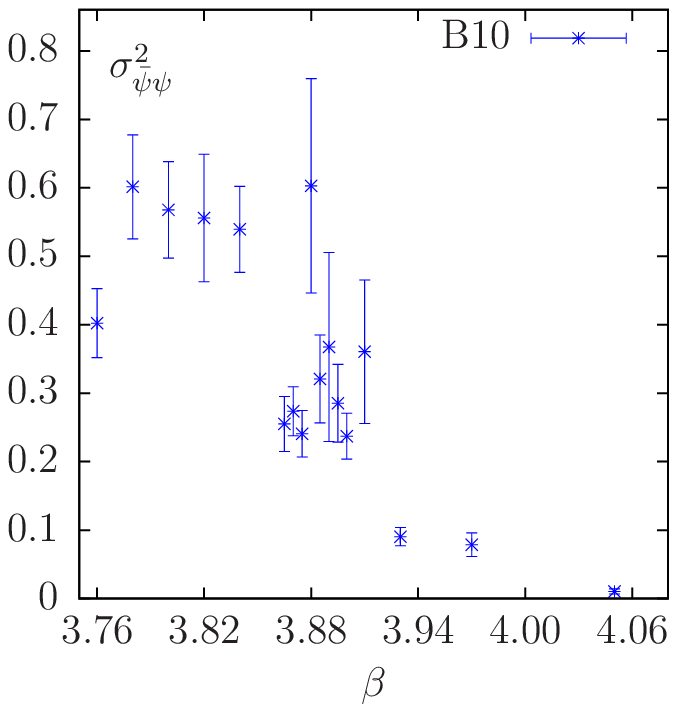} \hfill
\centering \hfill
\includegraphics[width=\graphicswidth]{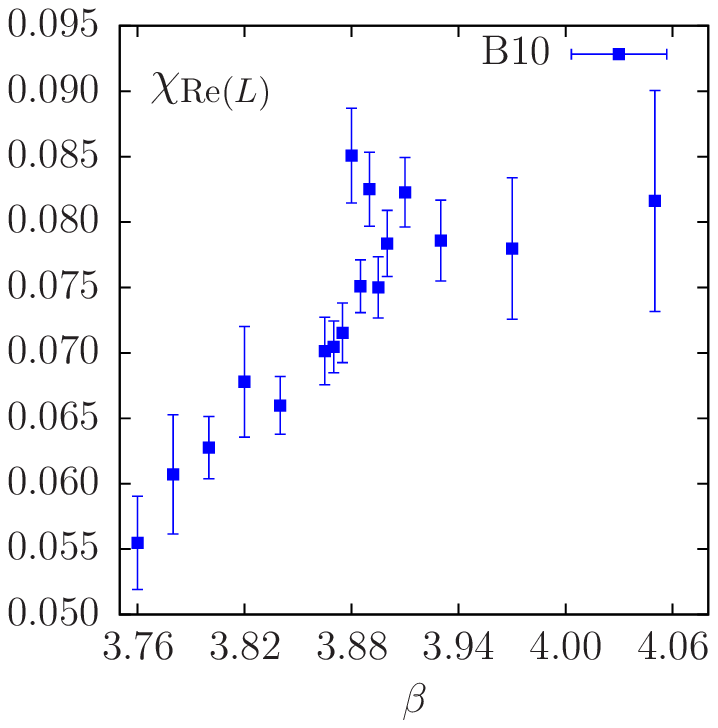}
\caption[]{$\sigma^2_{\pbp}$ (left) and susceptibility of \Real(L)
(right), both for run B10.}
\label{tc_B10}
\end{figure*}
%-------------------------------------------------------------------------

In \Figs \ref{tc_A12} - \ref{tc_B10} we show our data for $\sigma^2_{\pbp}$
in accordance with \Eq{eq:sigma2} and the susceptibility of the
real part of the Polyakov loop (\Eq{eq:polloop}).
We quite clearly see maxima for $\sigma^2_{\pbp}$ in all cases, whereas for 
the Polyakov loop susceptibility we find only an onset of certain shoulders 
for the ensembles A12, B12 and B10. At the higher pion mass case C12, where
we restricted ourselves to smaller statistics, there seems to appear 
a maximum also for the Polyakov susceptibility.
 
In order to estimate the pseudo-critical $\beta_c$ for chiral transition
we have modeled  the data for $\sigma^2_{\pbp}$ with a Gaussian
\begin{equation}
c + a \exp\left( - \frac{(\beta-\beta_{c})^2}{\sigma^2} \right)\;.
\label{eq:gau}
\end{equation}

The results for the corresponding pseudo-critical chiral transition temperature 
$T_\chi$ are collected in \Tab{tab:chiralresults}.

%---------------------------------------------------------------------------
\begin{table}
\centering
\begin{ruledtabular}
\begin{tabular}{lrrrr}
\textsc{Run} & $\Nt$ & \multicolumn{1}{c}{$\beta_c$} & $T_\chi$\,(MeV) & 
                       \multicolumn{1}{c}{$r_0T_\chi$}\\
\hline
\hline
 A12 & 12 & 3.89(3) & 202(7)       & 0.437(18) \\
 B12 & 12 & 3.93(2) & 217(5)       & 0.473(10) \\
 C12 & 12 & 3.97(3) & 229(5)       & 0.500(14) 
\end{tabular}
\end{ruledtabular}
\caption{List of pseudo-critical points for the chiral transition $T_\chi$.}
\label{tab:chiralresults}
\end{table}
%----------------------------------------------------------------------------

\begin{figure*}
\centering
\includegraphics[width=\graphicswidth,angle=-90]{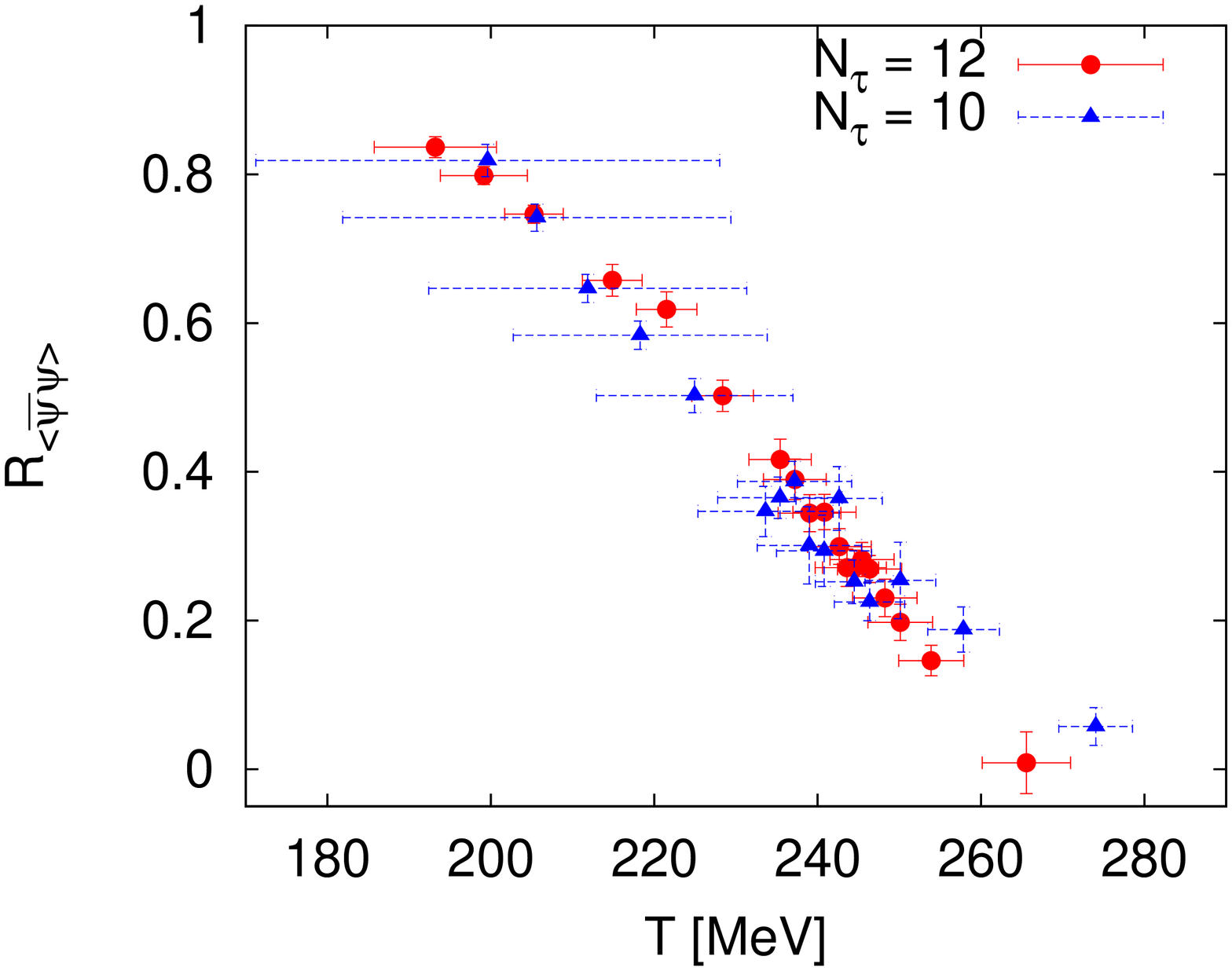}\hfill
\centering
\includegraphics[width=\graphicswidth,angle=-90]{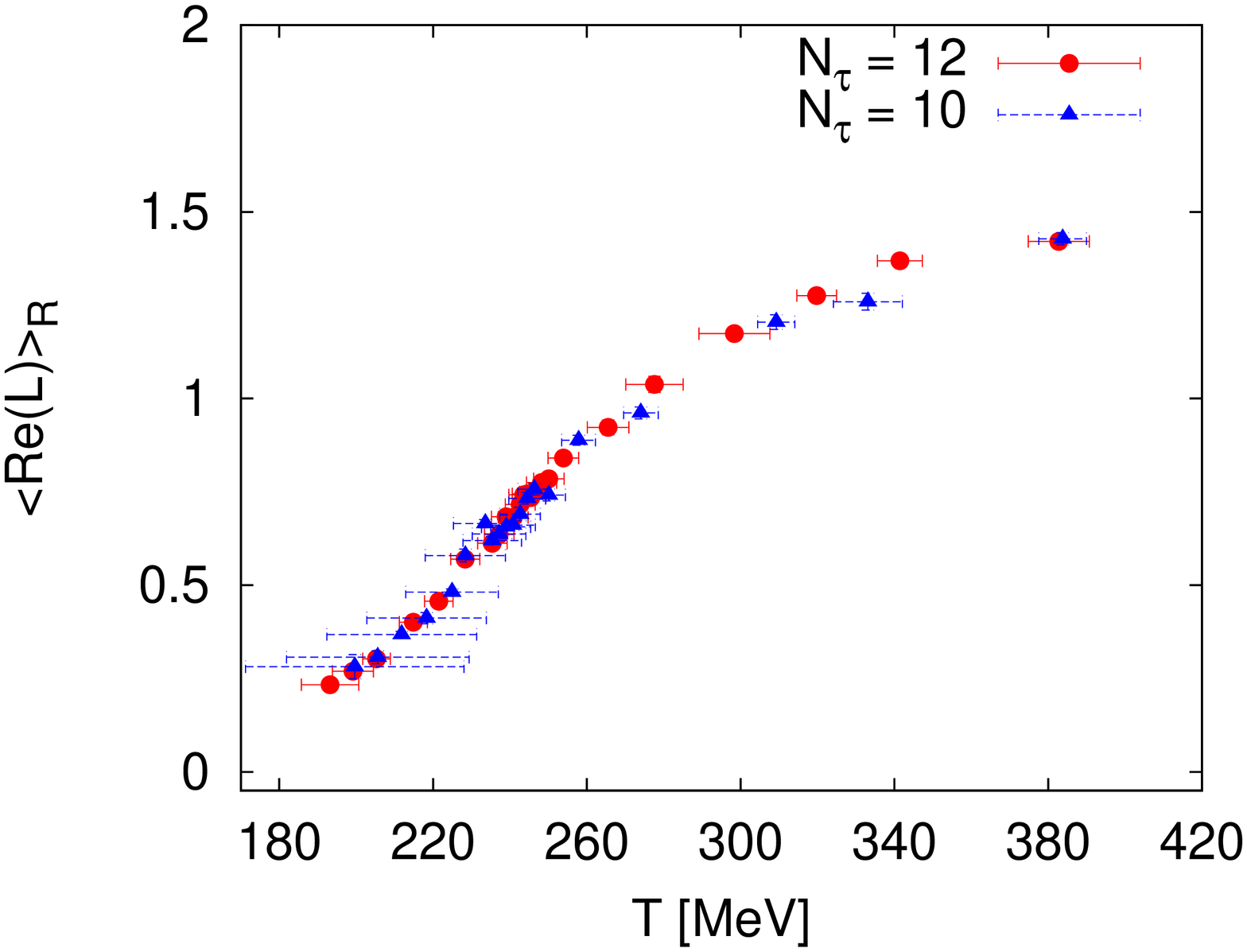}
\caption[]{
Ratio $ R_{\pbpvac}$ acc. to \Eq{eq:Rratio} (left) 
and renormalized Polyakov loop
$\vev{\Real(L)}_R$ (right), both for runs B12 and B10.}
\label{fig:renorm}
\end{figure*}

In \Fig{fig:renorm} we show the renormalized chiral condensate ratio 
$ R_{\pbpvac}$ and the renormalized Polyakov loop $\vev{\Real(L)}$ for 
the ensembles B12 and B10.  
The large error bars for the $T$-values in case of the B10 ensemble reflect the
uncertainty in the scale setting.
%----------------------------------------------------------------------------
%----------------------------------------------------------------------------

By determining the inflection point of the renormalized Polyakov loop $\vev{\Real(L)}$
we were able to estimate the deconfinement temperatures $T_\text{deconf}$ for the 
ensembles B12 and C12, see \Tab{tab:deconfresults}. 

%-------------------------------------------------------------------------------------------
\begin{table}
\centering
\begin{ruledtabular}
\begin{tabular}{lrrrr}
\textsc{Run} & $\Nt$ & \multicolumn{1}{c}{$\beta_c$} & $T_\text{deconf}$\,(MeV) & 
                       \multicolumn{1}{c}{$r_0T_\text{deconf}$}\\
\hline
\hline
 B12 & 12 &  4.027(14) &  249(5)  & 0.546(13) \\
 C12 & 12 &  4.050(15) &  258(5)  & 0.565(14)
\end{tabular}
\end{ruledtabular}
\caption{List of pseudo-critical points for the deconfinement transition $T_\text{deconf}$.}
\label{tab:deconfresults}
\end{table}
%-------------------------------------------------------------------------------------------
We cleary see that $~T_\text{deconf} > T_\chi~$ for both higher pion masses. 
This corresponds to the observation reported in 
\cite{Aoki:2006br}.

From weak coupling analyses (valid at high temperature) it is known that the leading order 
$a^2$-scaling towards the continuum limit might not set in before 
$\Nt\gtrsim16$~\cite{Philipsen:2008gq}. Therefore, discretization effects as a major source 
of systematical errors need to be thoroughly checked. Since the runs B10 and B12 share a common 
pion mass and differ only by $\Nt$, they can be used in order to assess the magnitude of cutoff 
effects. As can be seen from \Fig{tc_B10} the quality of $\sigma^2_{\pbp}$ for the B10 ensemble 
is not yet precise enough to allow for a Gaussian fit. 
The available data however suggests a maximum at around $\beta \sim 3.82$ which corresponds
to a temperature $T\sim218 \mathrm{~MeV}$ and agrees with $T_\chi$ at $N_\tau=12$.
Moreover the renormalized Polyakov loop and the renormalized chiral condensate
(\Fig{fig:renorm}) agree within errors for B10 and B12 indicating small cutoff effects.

\section{Towards the chiral limit} \label{sec:chiral}
%-----------------------------------------------------

As indicated in the Introduction, the main interest in the $N_f=2$ thermal transition lies 
in its chiral limit, for which one would like to unequivocally determine the order of the 
phase transition. The chiral condensate $\langle \bar{\psi}\psi \rangle$ then is an order 
parameter corresponding to the magnetization in an appropriate spin model of the same 
universality class. Finite quark (and therefore pion) masses break the chiral symmetry 
explicitly, thus corresponding to an external field. 
Provided the $N_f=2$ chiral limit features a second order transition and belongs to the 
$O(4)$ universality class, one may extrapolate finite mass simulations using universal 
scaling relations, which hold within some scaling region around the critical phase 
transition~\cite{Karsch:1994hm,Iwasaki:1996ya}. A priori it is not known how far into the 
massive region scaling extends, i.e.~one can merely test consistency of the data with scaling. 
A further difficulty is that chiral symmetry is broken explicitly for Wilson fermions at finite 
lattice spacing, even in the massless case. Any universal behavior for these types of fermions 
thus corresponds to continuum scaling, which can only be observed once discretization errors 
are sufficiently small. Finally, the scaling relations we employ here are valid
in the thermodynamic limit. Dedicated finite-size scaling analyses are required to establish
the appropriate lattice sizes, but this is beyond the scope of the present study. Again, 
we assume our lattices to be sufficiently large and test for consistency with scaling.

\begin{figure}
\centering
\includegraphics[width=1.3\graphicswidth,angle=0]{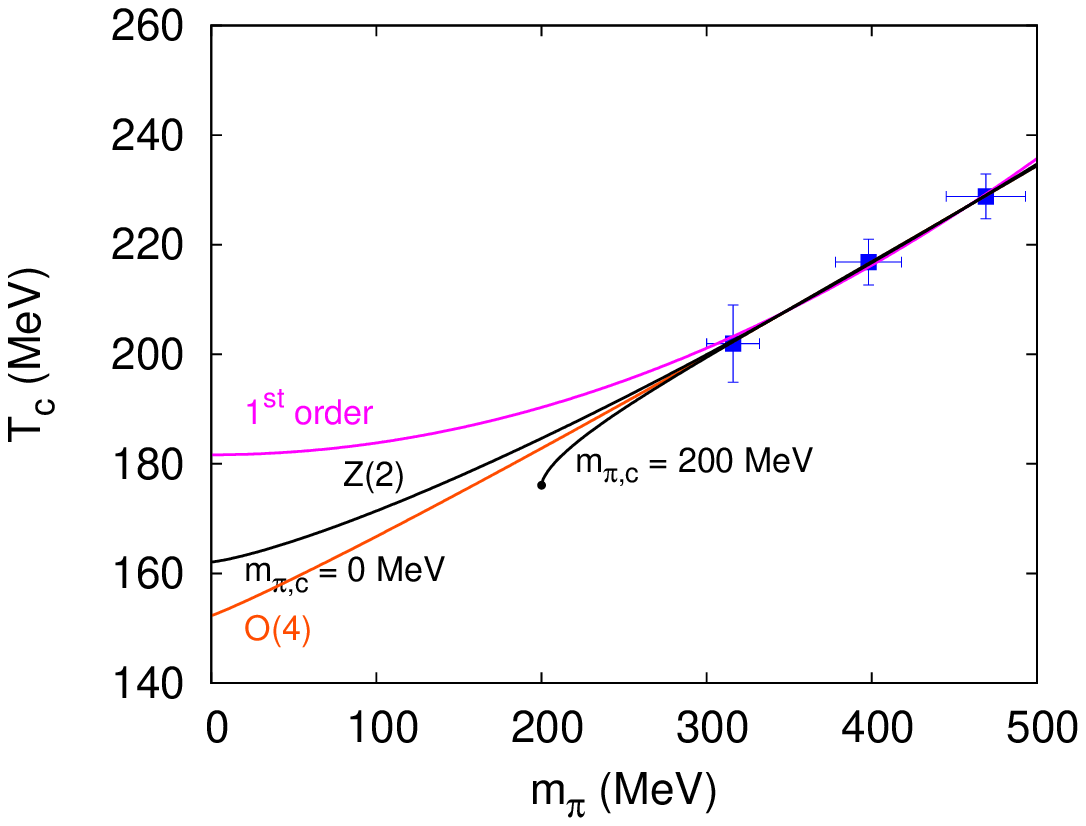}
\caption{Chiral extrapolation for 
$T_\chi(m_\pi)$
for various scenarios as explained in the text.
}
\label{fig:chiralfit:plot}
\end{figure}

\begin{figure}
\centering
\includegraphics[width=\graphicswidth]{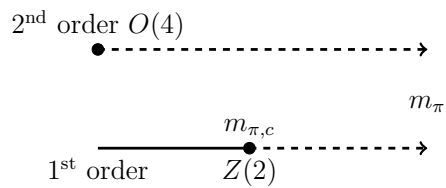}
\caption{Illustration of possible scenarios for the $N_f=2$ chiral limit.}
\label{fig:chiralfit:scenarios}
\end{figure}

We begin by attempting a fit of $T_\chi(m_\pi)$ to the scaling 
form ~\cite{Karsch:1993tv,Iwasaki:1996ya}
\begin{equation}
T_\chi(m_\pi) = T_\chi(0) + A \cdot m_\pi^{2/(\tilde\beta\delta)}\;, \label{eq:chiralfit}
\end{equation}
where we have dressed the critical exponent $\tilde\beta$ with a tilde in order to 
distinguish it from the lattice coupling. The ``external field'' in this case is the 
quark mass specified by the mass parameter $a\mu_0$, which in turn is connected to the 
pion mass in LO\chpt{} via $m_\pi^2  \sim \mu_0$. Thus, it is important to keep the pion 
mass small for two reasons, the validity of both the scaling window and the LO of \chpt. 
While there is good reason to expect that our pion masses are sufficiently small for 
the latter \cite{Baron:2009wt}, the size of the scaling region remains unknown at present. 
Unfortunately, we do not have sufficiently many data points or sufficiently small errors
in order to determine the exponents, but fix the exponents and fit $A$ and $T_c(0)$ only.
For $O(4)$ we have $2/(\tilde\beta\delta) = 1.08 $ and the resulting extrapolation is 
shown in \Fig{fig:chiralfit:plot}, giving a chiral critical temperature 
$T_\chi(m_\pi=0)=152(26)$ MeV.
It is now interesting to ask whether $O(4)$ scaling can be discriminated from other
behavior. As discussed earlier, the alternative scenario is a first order phase 
transition in the chiral limit. Often in the literature the same scaling relation is 
tested by merely changing to ``first order exponents'' 
($2/(\tilde\beta\delta) = 2$)~\cite{Cheng:2006qk,Bornyakov:2009qh}. Doing so leads 
to an extrapolation with somewhat larger $T_\chi(m_\pi=0)=182(14)$ MeV. 
However, it is unclear to us whether the scaling relation is applicable in this case.
Firstly, for a first order phase transition there is no diverging correlation length. 
Approaching $T_\chi$ in the infinite-volume limit from above and below proceeds in 
different phases, with finite correlation length in each. 
Hence, there is no scaling and no universality in the sense of second order transitions
(in particular $\tilde\beta=0$ and $\delta=\infty$ separately). 
The ``critical exponents'' usually associated with first order transitions specify the 
approach of the thermodynamic limit in finite-size scaling analyses, but do not apply 
to the relation (\ref{eq:chiralfit}) in the thermodynamic limit (for a detailed discussion 
of scaling for first order phase transitions, see \cite{Fisher:1982xt}).
Secondly, if the chiral limit indeed features a first order phase transition, it will 
weaken with finite quark masses until it vanishes in a $Z(2)$ critical endpoint. 
\Fig{fig:chiralfit:scenarios} shows the two possible scenarios. However, this means 
that coming from the crossover region at larger quark masses, an extrapolation to 
the chiral limit is never exact, as it would pass through a singularity at the 
critical point. Rather, the approach of this singularity will again be characterized 
by scaling, this time in the $Z(2)$ universality class. In this case we may use again 
the relation (\ref{eq:chiralfit}), but with a finite critical pion mass marking the 
critical point, $m_\pi^2\rightarrow (m_\pi^2-m_{\pi,c}^2)$.  We have attempted such 
extrapolations also.
Our data are not sufficient to constrain $m_{\pi,c}$. Therefore, \Fig{fig:chiralfit:plot} 
shows two extrapolations, one with $m_{\pi,c}\approx 0$ and another with 
$m_{\pi,c}\approx 200$ MeV. As the figure illustrates, our extrapolations alone 
cannot yet discriminate between the first order and second order scenarios. 
This would require drastically smaller pion masses, lower than the physical value even. 
Nevertheless, utilising knowledge about $T_c$ from other simulations we still obtain
a tendency. The fit assuming a first order scenario leads to a critical temperature 
which is somewhat larger than expected from other investigations~\cite{Kanaya:2010qj}. 
Of course, those extrapolations are likewise valid only in the  $O(4)$ scenario, 
so again this is merely a consistency test. 

%-------------------------------------------------------------------------

%-------------------------------------------------------------------------

\begin{figure}
\centering
\includegraphics[width=\graphicswidth,angle=-90]{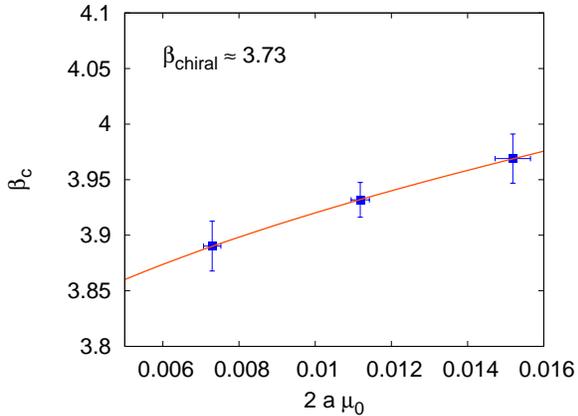} \hfill
\caption{Critical couplings 
$\beta_c$ as function of the external field $h$.}
\label{fig:betachiral}
\end{figure}

For a fixed \Nt{}, assumed to be large enough so as to be sufficiently close 
to the continuum, it is also possible to obtain the chiral critical $\beta$ 
by means of the scaling relation~\cite{Karsch:1993tv,Iwasaki:1996ya}
\begin{equation}
\beta_c(h) = \beta_\text{chiral} + B\cdot h^{1/(\tilde\beta\delta)}\;, \qquad
h=2a\mu_0 
\label{eq:betachiral}
\end{equation}
with $1/(\tilde\beta\delta) = 0.537$ corresponding to $O(4)$ exponents. 
For $\Nt=12$ our estimates for $\beta_c$ are shown in \Fig{fig:betachiral}
and can be extrapolated in this manner. 

Consistent fits have been found taking all three points from A12 to C12 into account. 
The result for the critical chiral $\beta$-value is
\begin{equation}
\beta_\text{chiral}(\Nt=12) = 3.73(9)\;.
\end{equation}
We have carried out the same fit but with the two lower pion mass values (A12 and B12)
only. It ended up with the same value.
%-----------------------------------------------------------------
%-----------------------------------------------------------------
This result corresponds to  $T_\chi(m_\pi=0) \approx 152(26)$\,MeV
where the error results from he scale setting. This number is in accord 
with our fits for $T_\chi(m_\pi)$ for a second order transition in the 
chiral limit. Note however, that the lattice spacing necessary to set 
the scale stems from an extrapolation to smaller values of $\beta$ than 
available from ETMC. This is reflected in the large uncertainty assigned 
to the temperatures. 

%-----------------------------------------------------------------------------------
\begin{figure}
\includegraphics[width=1.3\graphicswidth]{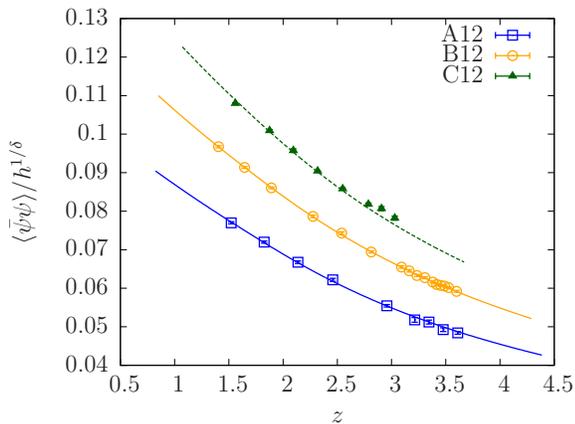}
\caption{Scaling for the bare \vev{\pbp} data at $\Nt=12$  as function of 
the scaling variable with modelling of scaling violations. 
The fit shown is for the combined A12 and B12 data (fit No 10 in \Tab{tab:scaling}).}
\label{fig:scaling}
\end{figure}
%------------------------------------------------------------------------------------

Next, the scaling of the magnetic equation of state can be
investigated, \Fig{fig:scaling}, where we follow previous 
studies~\cite{Iwasaki:1996ya,Ejiri:2009ac}
\begin{equation}
\vev{\pbp} = h^{1/\delta} c f(d\;\tau/h^{1/(\tilde\beta\delta)})\;,
\end{equation}
with
\begin{equation}
 \tau=\beta-\beta_\text{chiral}\;.
\end{equation}
The functional form of the scaling function $f$ for the $O(4)$ case is 
known~\cite{Toussaint:1996qr,Engels:1999wf}. Since we do not know the 
correct normalization for $\tau$ and $h$ with respect to QCD, 
we are left with two free parameters, $c$ and $d$, that have to be 
fitted. We perform the fits in the $\beta$-intervals from $\beta=3.83 (3.85, 3.89)$ 
to $\beta = 3.97 (4.03, 4.04)$ for A12 (B12, C12), respectively. The
fit results are collected in \Tab{tab:scaling}.

A fit for the line A12 works quite well, $\chi^2/\text{dof}=0.43$, but
giving $\beta_\text{chiral}=3.57(4)$, which is smaller than 
the value estimated above by applying \eq{eq:betachiral}. 
In general we observe an increase of $\chi^2$ and a decrease of 
$\beta_\text{chiral}$ with increasing mass. Indeed, B12 yields 
$\beta_\text{chiral}=3.40(5)$, which would correspond to a much too low
critical temperature below $100$ MeV, while C12 gives even smaller values 
with larger $\chi^2$. Thus, the fit seems to account for scaling 
violations due to large mass by decreasing $\beta_\text{chiral}$. 

However, scaling violations due to the quark mass can be taken into 
account by an ansatz including corrections~\cite{Ejiri:2009ac},
\begin{equation}
\vev{\pbp} = h^{1/\delta} cf(d\;\tau/h^{1/(\tilde\beta\delta)}) + 
             a_t \tau h + b_1 h + b_3 h^3 +\ldots\;. 
\label{eq:scaleviolations}
\end{equation}
We have fitted our data in numerous ways by taking into account one, two 
or even three violation terms.
Joint fits to the A12 + B12 ensembles are feasible in all three combinations, 
giving a $\beta_\text{chiral}$ the more consistent with the previous 
determination the more violation terms are included (see \Tab{tab:scaling}).  
In \Fig{fig:scaling} we show a combined fit to A12 and B12 fixing 
$\beta_\text{chiral}=3.73$ from our independent determination with 
$\chi^2/\text{dof}=0.63$.
Note that these fits with the two lower order violation terms are not 
able to include the C12 data with the requirement of a reasonable value of 
$\chi^2/\text{dof}$. 
However, if we include the next higher violation term $b_3 h^3$ in
the combined fit to A12, B12 and C12 we obtain 
an acceptable $\chi^2/\text{~dof} = 1.8$, see the the last line of 
\Tab{tab:scaling}. We observe that in this case the fit even 
prefers a value for $\beta_\text{chiral}$ compatible with the one from 
the analysis based on \eq{eq:betachiral}.

Since we are in a range of the scaling variable $\tau/h^{1/(\tilde\beta\delta)}$ 
where the scaling function is rather flat, judgement on whether there are 
additional violations of the $O(4)$ behavior or not is difficult. Repeating 
this exercise for the first order scenario with endpoint does not give 
further insight as the combinations of exponents are very close, 
$1/(\tilde\beta\delta)=0.537, 0.638$ and $1/\delta=0.21,0.20$  
for $O(4)$ and $Z(2)$, respectively. Therefore, our data are
consistent with the $O(4)$ scenario, but do not rule out the possibility 
of the first order case. This would require drastically smaller pion 
masses combined with finite-size studies, as the window for chiral scaling 
appears to set in for $m_\pi \ll 300$ MeV.

%---------------------------------------------------------------------------------------------------------
\begin{table*}
\centering
\begin{ruledtabular}
\begin{tabular}{ccrrrrrrr}
\textsc{No}& \textsc{Data} & $\beta_\text{chiral}$ & $c$ & $d$ & $a_t$ & $b_1$ & $b_3$ & $\chi^2$/dof \\
\hline \hline

\hline
1 & A12            &3.57(4)        &0.14(2)   &0.367(7)  &\textbf{0}  &\textbf{0} &\textbf{0}  &0.43     \\
2 & B12            &3.40(5)        &0.22(4)   &0.36(2)   &\textbf{0}  &\textbf{0} &\textbf{0}  &0.64     \\
3 & C12	           &3.12(2)	   &0.42(3)   &0.39(2)   &\textbf{0}  &\textbf{0} &\textbf{0}  &2.42     \\
4 & A12 + B12      &3.368(6)       &0.257(6)  &0.383(5)  &\textbf{0}  &\textbf{0} &\textbf{0}  &3.31     \\
\hline
5 & A12 + B12      &3.48(2)        &0.225(6)  &0.48(2)   &0.7(1)      &\textbf{0} &\textbf{0}  &2.2      \\
6 & A12 + B12      &3.57(2)        &0.152(7)  &0.53(2)   &\textbf{0}  &0.90(6)    &\textbf{0}  &1.75     \\
7 & A12 + B12      &3.82(4)        &0.028(9)  &1.1(2)    &-2.2(2)     &2.49(8)    &\textbf{0}  &0.42     \\
\hline
8 & A12 + B12	   &\textbf{3.73}  &0.1279(8) &0.825(8)  &4.01(4)     &\textbf{0} &\textbf{0}  &76       \\
9 & A12 + B12      &\textbf{3.73}  &0.0759(7) &0.81(2)   &\textbf{0}  &1.61(2)    &\textbf{0}  &7.2      \\
10& A12 + B12	   &\textbf{3.73}  &0.053(2)  &0.74(2)   &-1.8(2)     &2.23(6)    &\textbf{0}  &0.63     \\
\hline
11&A12 + B12 + C12 &3.76(2)        &0.047(6)  &0.83(6)   &-1.5(2)     &2.20(6)    &50(11)      &1.8      \\
\end{tabular}
\end{ruledtabular}
\caption{Fit results based on \eq{eq:scaleviolations} for several 
combinations of our data sets and fit parameters. Numbers in bold face 
have been fixed before fitting. The fit shown in \Fig{fig:scaling} 
corresponds to line No 10.}
\label{tab:scaling}
\end{table*}
%-----------------------------------------------------------------------------------------------------------

\section{Conclusions} \label{sec:concl}
%--------------------------------------

We have presented a (revised) first
investigation of the two-flavor thermal QCD transition 
with maximally twisted mass fermions. Our results are compatible with existing 
work although, of course, staggered investigations are much more 
advanced~\cite{Aoki:2009sc,Borsanyi:2010bp,Borsanyi:2010cj,Cheng:2009zi}. 
The quality of our signals is comparable to recent results with clover improved 
Wilson fermions~\cite{Bornyakov:2009qh,Bornyakov:2011yb}. 

For three pion masses in the range  $300\,\text{MeV}<m_\pi<500\,\text{MeV}$ we 
have determined pseudo-critical temperatures for the crossover from the 
hadronic regime to the quark gluon plasma. The pseudo-critical temperatures 
- extracted for the two higher mass values - from observables related to chiral and 
deconfinement transitions, respectively, turned out to be different. 
Discretization effects in $T_c$ appeared to be small for our lattice spacings, 
$a<0.09$ fm.
 
We have restricted ourselves to pion masses $<500$\,MeV in order to assure 
the validity of LO\chpt{} as well as the scaling forms in order to extrapolate 
to the chiral limit. Assuming the scaling forms appropriate for different 
universality classes, such extrapolations gave critical temperatures  
in the range $T_c\sim 140-200$ MeV consistent with other studies. 
However, detailed fitting analyses demonstrated that the second order $O(4)$ 
scaling regime is not yet reached. Scaling violations could be accomodated by 
leading order corrections due to finite-mass effects up to $m_\pi\sim 400$ MeV, 
while heavier masses violate even those corrections. By including higher
order violation effects reasonable fits could be achieved with $\beta_c$
values consistent with the other determinations.

We find that truly 
distinguishing between the different universality classes and thus ruling 
out a first order scenario will require much smaller pion masses, 
$m_\pi\lsim m_\pi^{phys}$ as well as finite-size scaling analyses.
We hope to address these issues in future investigations.

\begin{acknowledgments}
We express our gratitude to Giancarlo Rossi for clarifying the issue of
renormalization in the twisted mass case.
Moreover, M.P.L.~and L.Z.~thank Roberto Frezzotti for useful discussions.
O.P.~ and L.Z.~are supported by DFG grant PH 158/3-1, C.P. by the 
German BMBF grant 06FY7100. F.B.~ and M.M.P.~acknowledge support by 
DFG GK 1504 and SFB/TR 9, respectively. 
We are grateful to the HLRN supercomputing centers Berlin and Hannover
and apeNext in Rome as well as the LOEWE-CSC in Frankfurt for computing 
resources. 
\end{acknowledgments}

%-------------------------------------------------------------------------------------------
\begin{table*}
\begin{ruledtabular}
\begin{minipage}[t]{.45\textwidth}
\vspace{0pt}
\centering
\begin{tabular}{rrrrcr}
\multicolumn{5}{c}{A12} \\
\hline\hline
\multicolumn{1}{c}{$\beta$} &$T [\mathrm{~MeV}]$& \textsc{stat} & \multicolumn{1}{c}{Re($L$)} 
&  \multicolumn{1}{c}{$\vev{\pbp}$}\\
\hline\hline
3.8400 & 187(10) &  3471 & $6.1(4)\cdot10^{-4}$  & 0.0284(1) \\
3.8600 & 193(8)  &  7114 & $6.7(3)\cdot10^{-4}$  & 0.0264(1) \\
3.8800 & 199(6)  &  3891 & $8.8(4)\cdot10^{-4}$  & 0.0243(1) \\
3.9000 & 205(4)  &  6666 & $9.8(4)\cdot10^{-4}$  & 0.0225(2) \\
3.9300 & 215(4)  &  3947 & $1.40(4)\cdot10^{-3}$ & 0.0199(1) \\
3.9450 & 220(4)  &  4839 & $1.60(5)\cdot10^{-3}$ & 0.0185(2) \\
3.9525 & 222(4)  &  5962 & $1.67(4)\cdot10^{-3}$ & 0.0183(2) \\
3.9600 & 225(4)  &  6112 & $1.86(5)\cdot10^{-3}$ & 0.0176(2) \\
3.9675 & 228(4)  &  7112 & $1.98(5)\cdot10^{-3}$ & 0.0172(2) \\
3.9750 & 230(4)  &  4505 & $2.06(6)\cdot10^{-3}$ & 0.0168(2) \\
3.9900 & 235(4)  &  4796 & $2.45(5)\cdot10^{-3}$ & 0.0158(2) \\
\end{tabular}
\vfill
\end{minipage}
\hfill
\begin{minipage}[t]{.45\textwidth}
\vspace{0pt}
\flushleft
\begin{tabular}{rrrcr}
\multicolumn{5}{c}{B12} \\
\hline\hline
\multicolumn{1}{c}{$\beta$} &$T [\mathrm{~MeV}]$&  \textsc{stat} & \multicolumn{1}{c}{Re($L$)} 
&  \multicolumn{1}{c}{$\vev{\pbp}$}\\
\hline\hline 
 3.8600  & 193(8) &  7198 & $  5.95(22)\cdot10^{-4}$ &  0.03916(12) \\
 3.8800  & 199(6) &  7883 & $  7.29(22)\cdot10^{-4}$ &  0.03677(10) \\
 3.9000  & 205(4) &  9568 & $  8.67(19)\cdot10^{-4}$ &  0.03444(09) \\
 3.9300  & 215(4) &  9204 & $  1.24(03)\cdot10^{-3}$ &  0.03122(13) \\
 3.9500  & 222(4) &  4788 & $  1.49(05)\cdot10^{-3}$ &  0.02932(14) \\
 3.9700  & 228(4) &  8387 & $  1.96(07)\cdot10^{-3}$ &  0.02724(10) \\
 3.9900  & 235(4) &  8968 & $  2.09(05)\cdot10^{-3}$ &  0.02557(13) \\
 3.9950  & 237(4) &  6486 & $  2.31(04)\cdot10^{-3}$ &  0.02515(13) \\
 4.0000  & 239(4) &  6298 & $  2.51(04)\cdot10^{-3}$ &  0.02464(11) \\
 4.0050  & 241(4) &  7353 & $  2.54(05)\cdot10^{-3}$ &  0.02438(10) \\
 4.0100  & 243(4) &  6403 & $  2.70(05)\cdot10^{-3}$ &  0.02391(10) \\
 4.0125  & 244(4) &  10139& $  2.81(04)\cdot10^{-3}$ &  0.02365(11) \\
 4.0150  & 245(4) &  8950 & $  2.84(04)\cdot10^{-3}$ &  0.02353(10) \\
 4.0175  & 245(4) &  11673& $  2.82(03)\cdot10^{-3}$ &  0.02346(09) \\
 4.0200  & 246(4) &  10003& $  2.88(04)\cdot10^{-3}$ &  0.02328(07) \\
 4.0250  & 248(4) &  9878 & $  3.02(04)\cdot10^{-3}$ &  0.02288(10) \\
 4.0300  & 250(4) &  5245 & $  3.14(05)\cdot10^{-3}$ &  0.02251(09) \\
 4.0400  & 254(4) &  5350 & $  3.43(05)\cdot10^{-3}$ &  0.02186(07) \\
 4.0700  & 266(6) &  1024 & $  4.00(08)\cdot10^{-3}$ &  0.02025(10) \\
 4.1000  & 278(8) &  7837 & $  4.83(09)\cdot10^{-3}$ &  0.01894(04) \\
 4.1500  & 298(10)&  4080 & $  6.17(07)\cdot10^{-3}$ &  0.01736(03) \\
 4.2000  & 320(6) &  4160 & $  7.57(08)\cdot10^{-3}$ &  0.01583(02) \\
 4.2500  & 341(6) &  4160 & $  9.17(07)\cdot10^{-3}$ &  0.01451(03) \\
 4.3500  & 383(8) &  4334 & $  1.21(01)\cdot10^{-2}$ &  0.01185(01) \\
\end{tabular}
\end{minipage}\\[.1cm]
\begin{minipage}[t]{.45\textwidth}
\vspace{0pt}
\centering
\begin{tabular}{rrrrcr}
\multicolumn{5}{c}{C12} \\
\hline\hline
\multicolumn{1}{c}{$\beta$} & $T [\mathrm{~MeV}]$& \textsc{stat} & \multicolumn{1}{c}{Re($L$)} 
&  \multicolumn{1}{c}{$\vev{\pbp}$}\\
\hline\hline
 3.9000 & 205(4) &  3050     & $  8.4(5)\cdot10^{-4} $ & 0.0465(2) \\
 3.9300 & 215(4) &  3101     & $  1.16(4)\cdot10^{-3}$ & 0.0431(2) \\
 3.9500 & 222(4) &  5822     & $  1.35(3)\cdot10^{-3}$ & 0.0407(2) \\
 3.9700 & 228(4) &  9179     & $  1.63(3)\cdot10^{-3}$ & 0.0379(2) \\
 3.9900 & 235(4) &  5151     & $  2.11(5)\cdot10^{-3}$ & 0.0360(2) \\
 4.0100 & 242(4) &  4640+5432& $  2.48(5)\cdot10^{-3}$ & 0.0341(2) \\
 4.0200 & 246(4) &  5120+3324& $  2.49(6)\cdot10^{-3}$ & 0.0336(3) \\
 4.0300 & 250(4) &  6240+3308& $  2.92(7)\cdot10^{-3}$ & 0.0325(3) \\
 4.0400 & 254(4) &  4080+3308& $  3.20(7)\cdot10^{-3}$ & 0.0315(3) \\
 4.0500 & 258(5) &  4640     & $  3.57(8)\cdot10^{-3}$ & 0.0306(2) \\
 4.0600 & 262(5) &  5523     & $  3.79(5)\cdot10^{-3}$ & 0.0296(1) \\
 4.0700 & 266(6) &  2790     & $  4.20(5)\cdot10^{-3}$ & 0.0288(1) \\
\end{tabular}
\end{minipage}
\hfill
\begin{minipage}[t]{.45\textwidth}
\vspace{0pt}
\flushleft
\begin{tabular}{rrrcr}
\multicolumn{5}{c}{B10} \\
\hline\hline
\multicolumn{1}{c}{$\beta$} &$T [\mathrm{~MeV}]$ & \textsc{stat} & \multicolumn{1}{c}{Re($L$)} 
&  \multicolumn{1}{c}{$\vev{\pbp}$}\\
\hline\hline
 3.7600 & 200(29) &  7760  & $ 1.57(07)\cdot10^{-3}$ & 0.05146(10)\\
 3.7800 & 206(24) &  3328  & $ 1.80(10)\cdot10^{-3}$ & 0.04769(15)\\
 3.8000 & 212(20) &  3097  & $ 2.25(06)\cdot10^{-3}$ & 0.04398(19)\\
 3.8200 & 218(16) &  3516  & $ 2.60(10)\cdot10^{-3}$ & 0.04091(17)\\
 3.8400 & 225(13) &  3279  & $ 3.19(08)\cdot10^{-3}$ & 0.03783(18)\\
 3.8650 & 228(11) &  3450  & $ 4.80(08)\cdot10^{-3}$ & 0.03364(19)\\
 3.8700 & 234(9)  &  5900  & $ 4.38(10)\cdot10^{-3}$ & 0.03357(14)\\
 3.8750 & 235(8)  &  3600  & $ 4.49(10)\cdot10^{-3}$ & 0.03351(18)\\
 3.8800 & 237(7)  &  8759  & $ 5.07(18)\cdot10^{-3}$ & 0.03233(44)\\
 3.8850 & 239(7)  &  6400  & $ 4.91(11)\cdot10^{-3}$ & 0.03222(40)\\
 3.8900 & 241(6)  &  7789  & $ 5.17(13)\cdot10^{-3}$ & 0.03258(36)\\
 3.8950 & 243(6)  &  4450  & $ 5.52(12)\cdot10^{-3}$ & 0.03143(24)\\
 3.9000 & 244(5)  &  5973  & $ 5.81(11)\cdot10^{-3}$ & 0.03101(20)\\
 3.9100 & 246(5)  &  7250  & $ 5.70(12)\cdot10^{-3}$ & 0.03085(39)\\
 3.9300 & 250(5)  &  8050  & $ 7.23(10)\cdot10^{-3}$ & 0.02967(14)\\
 3.9700 & 258(5)  &  7276  & $ 8.42(13)\cdot10^{-3}$ & 0.02465(07)\\
 4.0500 & 274(5)  &  8716  & $ 1.24(2)\cdot10^{-2} $ & 0.02060(03)\\
 4.1000 & 309(5)  &  1517  & $ 1.44(3)\cdot10^{-2} $ & 0.01873(04)\\
 4.2000 & 333(9)  &  4131  & $ 2.00(2)\cdot10^{-2} $ & 0.01564(01)\\
\end{tabular}
\end{minipage}
\caption{Statistics for gauge observables from our simulations as 
well as expectation values of Re($L$) and \vev{\pbp}. 
Note that the trajectory length differs between the runs. 
On the apeNEXT (B10, C12 except for $\beta=4.06$) $\tau=0.5$, 
on the HLRN (A12, B12, $\beta=4.06$ of C12) $\tau=1$.}
\label{tab:hmcstat}
\end{ruledtabular}
\end{table*}
%----------------------------------------------------------------------------------------

%\bibliography{biblio}

\end{document}